\documentclass[a4paper, onecolumn, 11pt, accepted=2024-10-25]{quantumarticle}
\pdfoutput=1
\usepackage[utf8]{inputenc}
\usepackage[english]{babel}
\usepackage[T1]{fontenc}
\usepackage{amsmath}
\usepackage{hyperref}
\usepackage{tikz}

\usepackage{amsthm}
\usepackage{latexsym}
\usepackage{amsfonts}
\usepackage{amssymb}
\usepackage{color}
\usepackage{bbm,dsfont}
\usepackage{graphicx}
\usepackage{enumerate}
\usepackage{comment}
\usepackage{ulem}

\usepackage{cleveref}

\usepackage[numbers]{natbib}

\newtheorem{proposition}{Proposition}
\newtheorem{theorem}{Theorem}
\newtheorem{lemma}{Lemma}

\theoremstyle{definition}

\newtheorem{example}{Example}
\newtheorem{definition}{Definition}

\newcommand{\hi}{\mathcal{H}} 
\newcommand{\lh}{\mathcal{L(H)}} 
\newcommand{\sh}{\mathcal{S(H)}} 
\newcommand{\shd}{\mathcal{S}(\mathcal{H}_d)} 
\newcommand{\eh}{\mathcal{E(H)}} 
\newcommand{\ip}[2]{\left\langle\,#1\,|\,#2\,\right\rangle} 
\newcommand{\kb}[2]{|#1\rangle\langle#2|} 
\newcommand{\tr}[1]{{\rm tr}\left[#1\right]} 
\newcommand{\id}{\text{\usefont{U}{dsrom}{m}{n}1}} 

\newcommand{\Mo}{\mathsf{M}}
\newcommand{\No}{\mathsf{N}}

\newcommand{\Mr}[2]{\mathcal{M}^{row}_{#1,#2}} 
 
\newcommand{\Mrall}{\mathcal{M}^{row}} 
\newcommand{\C}[3]{\mathcal{C}_{#1,#2}(#3)} 
\newcommand{\Call}[1]{\mathcal{C}(#1)} 
\newcommand{\Calle}[1]{\mathcal{C}^{\usim}(#1)} 
\newcommand{\uleq}{\preceq} 
\newcommand{\usim}{\simeq} 
\newcommand{\nuleq}{\npreceq}

\newcommand{\state}{\mathcal{S}} 
\newcommand{\Qd}{\mathcal{Q}_d} 
\newcommand{\Cd}{\mathcal{C}_d} 

\newcommand{\rank}[1]{\mathrm{rank}(#1)} 
\newcommand{\nrank}[1]{\mathrm{rank}_+(#1)} 
\newcommand{\psdrank}[1]{\mathrm{rank}_{psd}({#1})} 
\newcommand{\lmax}[1]{\lambda_{max}(#1)} 
\newcommand{\lmaxx}{\lambda_{max}} 
\newcommand{\lmaxw}{\lmaxx}
\newcommand{\lmin}[1]{\lambda_{min}(#1)} 
\newcommand{\lminw}{\lambda_{min}} 

\begin{document}

\title{Maximal Elements of Quantum Communication}

\author{Teiko Heinosaari}
\affiliation{Quantum Information and Computation, Faculty of Information Technology, University of Jyväskylä, 40014, Finland}
\email{teiko.heinosaari@jyu.fi}
\orcid{0000-0003-2405-5439}
\author{Oskari Kerppo}
\affiliation{Quantum Information and Computation, Faculty of Information Technology, University of Jyväskylä, 40014, Finland}
\email{oskari.e.o.kerppo@jyu.fi}
\orcid{0000-0002-3886-3157}

\maketitle

\begin{abstract}
A prepare-and-measure scenario is naturally described by a communication matrix that collects all conditional outcome probabilities of the scenario into a row-stochastic matrix. The set of all possible communication matrices is partially ordered via the possibility to transform one matrix to another by pre- and post-processings. By considering maximal elements in this preorder for a subset of matrices implementable in a given theory, it becomes possible to identify communication matrices of maximum utility, i.e., matrices that are not majorized by any other matrices in the theory. The identity matrix of an appropriate size is the greatest element in classical theories, while the maximal elements in quantum theory have remained unknown. We completely characterize the maximal elements in quantum theory, thereby revealing the essential structure of the set of quantum communication matrices. In particular, we show that the identity matrix is the only maximal element in quantum theory but, as opposed to a classical theory, it is not the greatest element. Quantum theory can hence be seen to be distinct from classical theory by the existence of incompatible communication matrices.
\end{abstract}

\section{Introduction}

One of the fundamental pursuits in quantum information science is to identify and investigate different kinds of communication scenarios, with the aim of finding differences to classical communication. Notable examples are super dense coding \cite{Bennett1992Communication}, quantum teleportation \cite{Bennett1993Teleporting}, and quantum key distribution \cite{Bennett1984Quantum}. 
Communication scenarios such as these are not only important in the study of entanglement and quantum theory in general, but they also pave the way for new applications in the field of quantum computation and cryptography.
Recently, there have been new findings where a communication advantage for quantum systems has been demonstrated. Classical cost of simulating correlations arising from entangled quantum states have been studied \cite{Massar2001Classical, Toner2003Communication}, while quantum communication advantages have been shown when the communicating parties share randomness as a resource \cite{Renner2023Classical, Ambainis2008Quantum, Guha2021quantumadvantage}.
However, a quantum advantage can also be demonstrated in other settings \cite{Gallego2010Device, Chaves2015Device, Saha2020Advantage, Buhrman2001Fingerprinting, Ambainis2002Dense} and without any extra resources \cite{Perry2015Communication, Emariau2022Quantum, Galvao2003Substituting, Heinosaari2023Simple}. Demonstrations such as these are of crucial importance, as it is not clear in which scenarios exactly are quantum systems better than their classical analogues. The same problem arises in quantum computation, too; while some algorithms are known that perform exponentially faster than existing algorithms on a classical computer, such as Shor's algorithm \cite{Shor1997Polynomial}, it is not known exactly for which computational problems an exponential speed-up is possible or if this speed-up is theoretically robust. In fact, a kind of tug-of-war competition seems to be going on between classical and quantum computing –- often when significant advances are made in quantum computing, classical computers are quick to catch up any lost ground on computing time \cite{Arute2019Quantum, Pan2022Simulation}.

In this article, we take a slightly different approach to previous works on the topic. 
We focus on the one-way communication scenario between two parties (Alice and Bob), where Alice prepares a classical or quantum system among a finite number of possible options and sends it to Bob, who then performs a fixed measurement on the state of the system. We assume that Alice and Bob do not share any extra resources such as shared randomness or entanglement. 
Their collaborative action is hence described by a communication matrix, which is simply a table of conditional probabilities for input-output pairs.
Given two communication matrices $A$ and $B$, we may ask if one of them is harder to implement than the other. 
Classical pre- and postprocessing of labels are described by row-stochastic matrices $L,R$, so that the relevant mathematical relation is $LAR=B$, which is called the ultraweak matrix majorization \cite{HeKe2019, HeKeLe2020}.
To present our results, we are going to expand on the theory of ultraweak matrix majorization and investigate the ordering of both quantum and classical communication matrices.

Rather than presenting a single communication scenario displaying a quantum advantage, we will give a complete and exhaustive mathematical characterization of one-way communication devices of ``maximum utility'' in the quantum case.
These correspond to maximal communication matrices in terms of ultraweak matrix majorization—ones that cannot be derived via pre- and postprocessing from any inequivalent communication matrix. In other words, the ultraweakly maximal elements in a given theory are the communication matrices of best utility in the sense that there do not exist any communication matrices that are stronger than them.
Our results bring a new intuitive understanding both mathematically and physically. In particular, our main result sheds new light on distinctions between classical and quantum communication devices.
Interestingly, this difference goes away if shared randomness is a freely available resource. Namely, as shown in \cite{Frenkel2015Classical}, the convex closures of the sets of classical and quantum communication matrices coincide. 
A convex mixture can be formed if there is a way to make coordinated random choices, which is exactly the meaning of shared randomness.
However, since the simulation of quantum communication matrices with classical communication matrices requires an unbounded amount of shared randomness \cite{Heinosaari2023Simple}, it is motivated to investigate also the situation without any shared randomness, as we do in the current work.

The rest of the article is organized as follows. In Sections \ref{section:communication-tasks} and \ref{section:ultraweak-matrix-majorization} we give a brief introduction to the theory of communication matrices and ultraweak matrix majorization. Section \ref{section:physically-motivated} analyzes ultraweak matrix majorization from the point of view of some physically motivated actions, and presents a novel result that is used in the main theorem. Finally, in Section \ref{section:maximal-communication-matrices} we give a complete characterization on which quantum communication devices are maximal in the preorder defined by the ultraweak matrix majorization. We close the article in Section \ref{section:discussion} by discussing the physical significance of our results.

\section{Communication tasks}\label{section:communication-tasks}

In a one-way communication scenario one party, Alice, attempts to send a message to another party, Bob. Usually the size of the message is taken to be limited. For instance, the message could consist of a fixed number of units of some specified physical communication medium. 
In the following we are considering an abstract operational description of communication scenarios. 
In particular, we typically describe the communication medium as a classical $d$-level system or a $d$-level quantum system, i.e., a quantum system associated with $d$-dimensional Hilbert space.

\begin{figure}
    \centering
    \includegraphics[width=\textwidth]{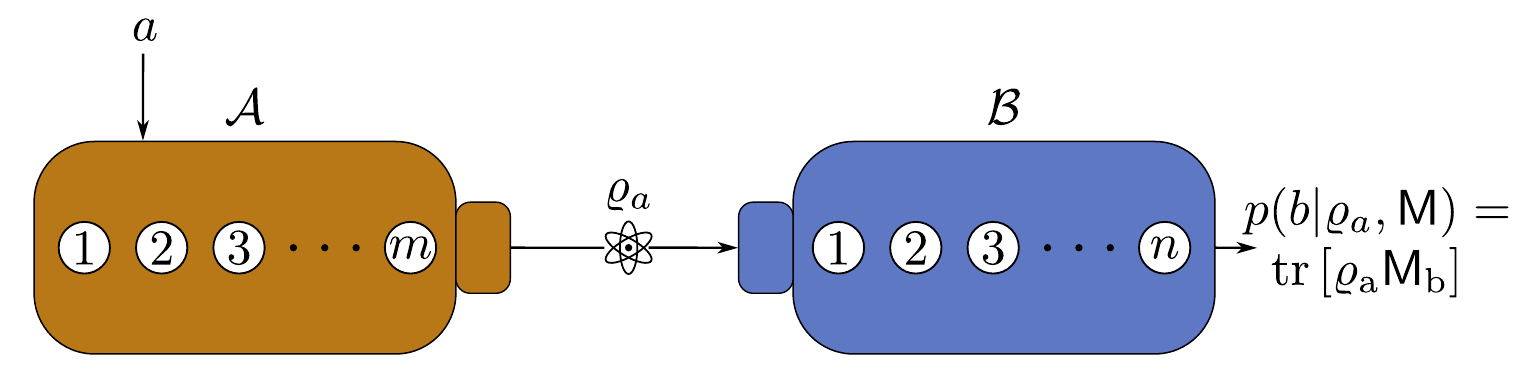}
    \caption{The basic quantum communication scenario. Alice chooses a label $a \in \{1, \dots , m\}$ and sends the state $\varrho_a$ to Bob. Bob performs a measurement of $\Mo=\{ \Mo_j \}_{j=1}^n$ and his device outputs an outcome $b$ with the probability $p(b |  \varrho_a,  \Mo) = \tr{\varrho_a \Mo_b}$.}
    \label{fig:basic-scenario}
\end{figure}

In the operational picture, depicted in Fig. \ref{fig:basic-scenario}, Alice operates a state preparation device with $m$ inputs.
This means that Alice has in her possession a device capable of preparing states of the communication medium with $m$ different labels. Some of the inputs of the device may lead to the same state being outputted. 
In each round of communication, Alice prepares one state from her device and sends it to Bob. 
Bob operates a measurement device with $n$ outputs. A measurement device is a device that takes as input a state and gives one of its outputs with a probability depending on the input state.

In the quantum setting, we describe Alice's state preparation device by a finite state ensemble $\{ \varrho_i \}_{i=1}^m$. Each of the states $\varrho_i$ in the state ensemble is a positive operator of unit trace. In general the quantum state space is described by $ \sh = \{ \varrho \in \lh \, | \, \varrho \geq 0 , \, \tr{\varrho}=1\}$, where $\hi$ is a complex separable Hilbert space, taken to be of finite dimension in this article, and $\lh$ is the set of bounded linear operators acting on $\hi$.

A quantum mechanical measurement device is described by a normalized \textit{positive operator measure}, or a POM for short (also called a POVM in the literature). More specifically, a measurement is described by a finite set of effect operators corresponding to different outcomes. The set of effects is defined as $\eh = \{ E \in \lh \, | \, 0 \leq E \leq \id \}$. A normalized POM $\Mo$ is then a finite collection of effects $\{ \Mo_i \}_{i=1}^n$ such that $\sum_i \Mo_i =\id$. In this work a POM is always taken to be normalized. Probabilities of different outcomes are given by the Born rule: \begin{align}\label{eq:born-rule}
    p(b |  \varrho,  \Mo) = \tr{\varrho \Mo_b},
\end{align}where $p(b |  \varrho,  \Mo)$ is the conditional probability of observing outcome $b$ when performing a measurement of $\Mo$ on state $\varrho$. Especially important to us will be POMs whose every effect is a rank-1 operator. Such POMs are simply called rank-1 POMs. A POM is sharp if its every effect is a projection. Rank-1 effects are also sometimes called indecomposable in the literature, as these effects cannot be written as positive linear combinations of other effects \cite{Kimura2010Distinguishability}.

Given a finite state ensemble $\{ \varrho_i \}_{i=1}^m$ and a POM $\Mo$ with $n$ outcomes, we can collect all the conditional outcome probabilities into a $m \times n$ row-stochastic matrix: \begin{align}
    C_{ab} = \tr{\varrho_a \Mo_b}.
\end{align}
Such a matrix $C$ is called a \textit{communication matrix} or \textit{communication channel}, and they arise naturally when considering outcome probabilities in a one-way communication scenario.

In a typical communication scenario, Alice and Bob are trying to accomplish some specific communication task. For instance, Alice and Bob's goal may be that Bob can unambiguously distinguish which state was sent by Alice, or they may try to minimize the error probability in distinguishing the state, leading to minimum error state discrimination. All communication tasks are hence described by some communication matrix, which defines the desired output probabilities for each state. As an example, communication matrices describing communication tasks of minimum error discrimination of three states with zero error and uniform antidistinguishability \cite{Bandyopadhyay2014Conclusive, Caves2002Conditions, Heinosaari2018Antidistinguishability} of three states are given as follows: \begin{align}\label{ex:communication-matrix-example}
    \id_3 = \begin{bmatrix}
    1 & 0 & 0 \\ 0 & 1 & 0 \\ 0 & 0 & 1
\end{bmatrix}, \, \qquad A_3 = \frac{1}{2}\begin{bmatrix}
    0 & 1 & 1 \\ 1 & 0 & 1  \\ 1 & 1 & 0
\end{bmatrix}.
\end{align}

An immediate question that arises when studying communication is which set of communication matrices is implementable in a given physical theory? Implementable here means that there exists a set of states and a measurement that implement the probabilities given by the communication matrix. If we denote a state space by $\state$, then the corresponding set of implementable communication matrices of size $m \times n$ is denoted by $\C{m}{n}{\state}$. The set of all such matrices is then $\Call{\state}=\cup_{m,n}\C{m}{n}{\state}$. The set of all $m$ by $n$ row-stochastic matrices is denoted by $\Mr{m}{n}$, or simply $\Mrall$ for the union of all row-stochastic matrices of finite size.

In the current investigation we focus on two special classes of state spaces. Firstly, we denote by $\Qd$ the state space in $d$-dimensional Hilbert space as a short-hand notation for $\shd$. In main focus will be the set $\Call{\Qd}$, the set of communication matrices implementable in $d$-dimensional quantum theory via the Born rule defined in Eq. \eqref{eq:born-rule}. Secondly, we denote a $d$-dimensional classical state space by $\Cd$. A set of communication matrices implementable by a classical $d$-level system is special in the sense that all communication matrices can be seen to be of a particular form \cite{HeKeLe2020}. Namely, if $C \in \Call{\Cd}$, then $C=L \id_d R$ for some row-stochastic matrices $L$ and $R$. For a given communication matrix $C$, the minimum value of $d$ for which there exists row-stochastic matrices $L$ and $R$ such that $C=L \id_d R$ is called the \textit{classical dimension} of the matrix $C$. The classical dimension is intimately related to the concept of nonnegative rank \cite{Cohen1993Nonnegative}. In fact, for communication matrices the two concepts coincide.

Analogously to the classical dimension, the \textit{quantum dimension} of a communication matrix $C \in \Mr{m}{n}$ is defined as the minimum value of $d$ for which there exists a state ensemble $\{ \varrho_i\}_{i=1}^m$ and a POM $\Mo$ with $n$ outcomes such that $C_{ab}=\tr{\varrho_a \Mo_b}$. For an arbitrary communication matrix $C$, the quantum dimension can be seen to be exactly given by the positive semidefinite rank (or psd rank \cite{Fawzi2015, Lee2017Some}) of $C$.

A second question regarding sets of communication matrices is if some communication matrices are harder to implement than others. As we will see in the next section, this question can be formulated in a precise mathematical way.

\section{Ultraweak matrix majorization}\label{section:ultraweak-matrix-majorization}

Given two communication matrices $A, B \in \Mrall$, we may ask if one of them is harder to implement than the other. 
Suppose there exist row-stochastic matrices $L,R$ such that $LAR=B$. 
It follows that $B$ cannot be any harder to implement than $A$. 
To see this, suppose $A$ has a quantum implementation with some state ensemble $\{ \varrho_i \}_{i=1}^m$ and a POM $\Mo$ with $n$ outcomes. 
Define new states $\varrho'$ by $\varrho'_k = \sum_{i=1}^m L_{ki}\varrho_i$ and a new POM by $\Mo'_k = \sum_{j=1}^n R_{jk}\Mo_j$. 
Then we get $$B_{ij} = (LAR)_{ij} = \sum_{k,l} L_{ik}A_{kl}R_{lj}=\sum_l R_{lj} \tr{(\sum_k L_{ik}\varrho_k)\Mo_l} = \tr{\varrho'_i \Mo'_j}.$$ 
That is, the matrix $B$ has an implementation by pre- and post-processing the states and measurement outcomes in $A$'s implementation. The pre- and post-processings are naturally given by the matrices $L$ and $R$. Whenever $LAR = B$ in the above sense, we say that the matrix $B$ is ultraweakly majorized by the matrix $A$.

\begin{definition}\label{def:ultraweak-matrix-majorization}
    Let $A,B \in \Mrall$. 
    If there exists row-stochastic matrices $L$ and $R$ such that $LAR=B$, we say that $B$ is ultraweakly majorized by $A$ and denote $ B \uleq A$. Whenever $A \uleq B$ and $B \uleq A$ we denote $A \usim B$ and say $A$ and $B$ are ultraweakly equivalent. If $A\nuleq B$ and $B \nuleq A$, we say that $A$ and $B$ are ultraweakly incomparable. Each matrix $A \in \Mrall$ defines an ultraweak equivalence class given by $[A] = \{ B \in \Mrall \, | \, B \usim A \}$.
\end{definition}

The ultraweak matrix majorization, along with monotone functions characterizing it, have been studied in detail in \cite{HeKe2019, HeKeLe2020}.
It is straightforward to see that the ultraweak matrix majorization is a preorder (i.e. reflexive and transitive) on the set of communication matrices.
In the usual way, this preorder determines a partial order in the set of all equivalence classes of communication matrices. 
We denote by $\Calle{\state}$ the set of all equivalence classes of matrices in $\Call{\state}$.

We recall the definition of a monotone function for ultraweak matrix majorization and introduce some useful monotones we are going to use. A thorough treatment can be found in \cite{HeKeLe2020}.

\begin{definition}
    A function $f: \Mrall \rightarrow \mathbb{R}$ is an ultraweak monotone function if $A \uleq B$ implies $f(A) \leq f(B)$ for all $A$ and $B$ in $\Mrall$.
\end{definition}

Some known ultraweak monotone functions include the following. For all matrices $A\in \Mrall$ the functions\begin{itemize}
    \item $\rank{A}$
    \item max monotone \cite{Matsumoto2018Information}: $\lmax{A}:= \sum_j \max_i(A_{ij})$
    \item min monotone \cite{HeKeLe2020}: $\lmin{A}:= -\sum_j \min_i(A_{ij})$  
    \item distinguishability monotone \cite{HeKeLe2020}: $\iota(A):= \max \{n \,|\, \id_n \uleq A \}$
    \item nonnegative rank \cite{Cohen1993Nonnegative}: $\nrank{A}$
    \item positive semidefinite rank \cite{Fawzi2015}: $\psdrank{A}$
\end{itemize} are ultraweak monotones. We recall that for all $A\in\Call{\Qd}$ we have $\lmax{A}, \iota(A) \leq d$, $\lmin{A} \leq 0$ and $\rank{A} \leq d^2$. For $A \in \Call{\Cd}$ we have $\lmax{A}, \iota(A) \leq d$, $\lmin{A} \leq 0$ and $\rank{A} \leq d$.

As mentioned before, the nonnegative and positive semidefinite ranks correspond to the optimal dimensions of the classical and quantum implementations of the matrix. The regular rank is connected to the linear dimension of the underlying operational theory. $\lmaxw$ is related to the optimal success probability in minimum error discrimination tasks (and the basic decoding theorem \cite{Schumacher2010Quantum}), while $\lminw$ has an operational interpretation relating to noise. The monotone $\iota(A)$ gives the exact number of distinct code words that are possible to send without error by the devices implementing $A$. In the present article we are going to leverage the monotones $\lmaxw$, $\lminw$ and $\iota$. Example \ref{ex:ultraweak-majorization} demonstrates the usefulness and application of the monotone $\lmaxw$. Example \ref{ex:incomparable} shows that there are incomparable communication matrices in quantum theory.

\begin{figure}
    \centering
    \includegraphics[width=\textwidth]{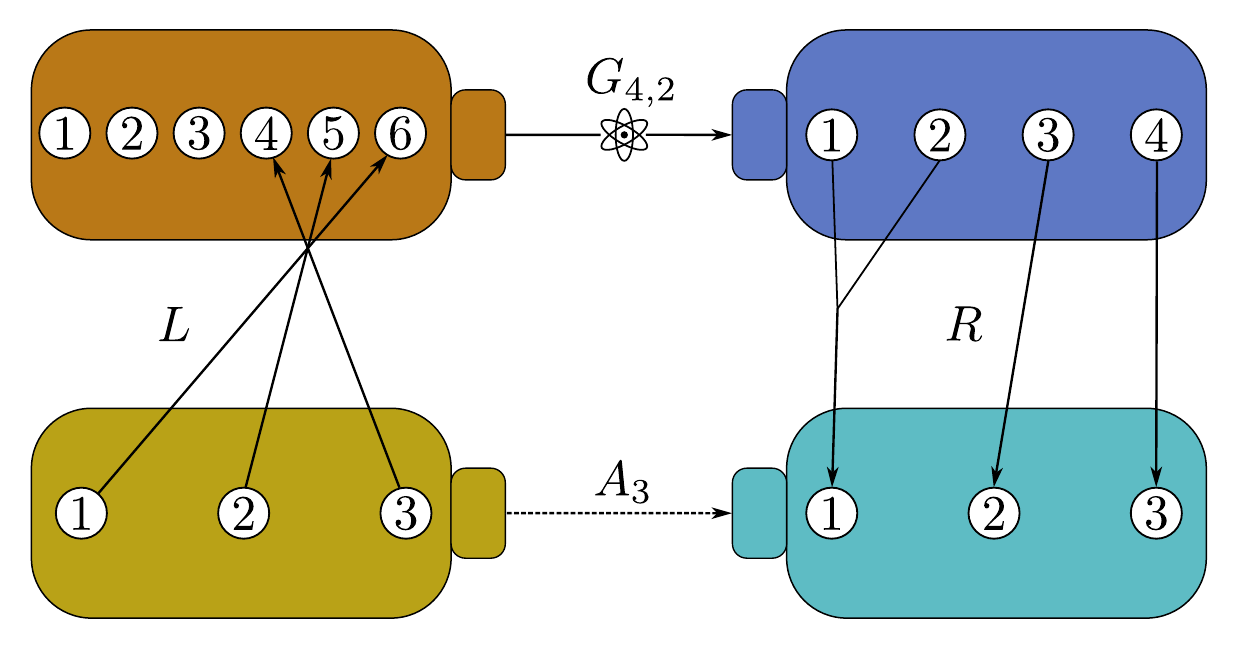}
    \caption{The communication matrix $G_{4,2}$ ultraweakly majorizes the communication matrix $A_3$. Thus $A_3$ has an implementation with the state preparation and measurement devices that implement $G_{4,2}$.}
    \label{fig:ultraweak-relation-example}
\end{figure}

\begin{example}\label{ex:ultraweak-majorization}
    Consider the following communication matrices: $$A_3 = \frac{1}{2}\begin{bmatrix}
    0 & 1 & 1 \\ 1 & 0 & 1 \\ 1 & 1 & 0
\end{bmatrix}, \, \quad G_{4,2} = \frac{1}{2}\begin{bmatrix}
    1 & 1 & 0 & 0 \\ 1 & 0 & 1 & 0  \\ 1 & 0 & 0 & 1 \\ 0 & 1 & 1 & 0 \\ 0 & 1 & 0 & 1 \\ 0 & 0 & 1 & 1
\end{bmatrix},
$$ 
and row-stochastic matrices
$$L = \begin{bmatrix}
    0 & 0 & 0 & 0 & 0 & 1 \\ 0 & 0 & 0 & 0 & 1 & 0 \\ 0 & 0 & 0 & 1 & 0 & 0 
\end{bmatrix} , \, \quad R = \begin{bmatrix}
        1 & 0 & 0 \\ 1 & 0 & 0 \\ 0 & 1 & 0 \\ 0 & 0 & 1 
    \end{bmatrix}.$$

We have $L G_{4,2} R = A_3$, hence $A_3 \uleq G_{4,2}$. However, we see that $\lmax{A_3}=\frac{3}{2} < 2 = \lmax{G_{4,2}}$. Therefore $\lmaxw$ witnesses that $G_{4,2} \npreceq A_3$, i.e., it is not possible to find $L, R \in \Mrall$ such that $G_{4,2} = LA_3R$. The relation $A_3 \uleq G_{4,2}$ is illustrated in Fig. \ref{fig:ultraweak-relation-example}.
\end{example}

\begin{example}\label{ex:incomparable}
As pointed out in \cite{HeKeLe2020}, $\Call{\Qd}$ contains communication matrices that are incomparable with $\id_d$.
For instance, let $A_n$ be the communication matrix that has $0$ in the diagonal and $\tfrac{1}{n-1}$ in all other entries. 
As explained in \cite[Example 4]{HeKeLe2020}, we have $A_{d+1}\in\Call{\Qd}$. 
Using the previously listed monotones we observe that $A_{d+1}$ and $\id_d$ are ultraweakly incomparable.
Namely, $\rank{\id_d}=d$ and $\lmax{\id_d}=d$, while $\rank{A_{d+1}}=d+1>d$ and $\lmax{A_{d+1}}=(d+1)/d <d$.  
\end{example}

As we explained earlier, all classical communication matrices belonging to $\Call{\Cd}$ are characterized by being ultraweakly majorized by $\id_d$. Thus the matrix $\id_d$ is the unique greatest element in classical $d$-dimensional theory, in the sense that $[\id_d]$ is the greatest element in the partially ordered set of equivalence classes in $\Call{\Cd}$. Motivated by this we make the following definition.

\begin{definition}\label{def:ultraweakly-maximal}
    Let $A \in \Call{\state}$.
    We say that $A$ is ultraweakly maximal in $\Call{\state}$ if $A \uleq B$ implies $A \usim B$ for all $B \in \Call{\state}$.
\end{definition}

The characterization of all maximal elements of the ultraweak preorder was stated as an open problem in \cite{Kerppo2023Quantum}. In the present work we will fully address this problem in the quantum case. Surprisingly, it turns out that quantum and classical theories are similar in that $\id_d$ is also the only maximal element in $\Call{\Qd}$. In order to see this, we will first give additional physical motivation to the ultraweak matrix majorization in the next section.

\section{Physically motivated pre- and post-processing actions}\label{section:physically-motivated}

For two communication matrices $A,B \in \Mrall$ it is straightforward to verify that $A$ and $B$ are ultraweakly equivalent in the following instances: \begin{itemize}
    \item[1)] $B$ is obtained from $A$ via permuting rows or columns
    \item[2)] $B$ is obtained from $A$ by adding a row that is a convex mixture of existing rows of $A$
    \item[3)] $B$ is obtained from $A$ by splitting a column according to some convex weights
\end{itemize}

All of these three actions have a clear operational meaning. \begin{itemize}
    \item[1)] Relabel states and measurement outcomes bijectively
    \item[2)] Add a mixture of existing states into the state ensemble implementing the communication matrix
    \item[3)] Split a measurement outcome into several outcomes probabilistically
\end{itemize}

A permutation of rows or columns is a reversible operation, but we also note that the operations in 2) and 3) can be naturally reversed: a new convex combination of rows can be added, but a convex combination of other rows can also be removed. Similarly columns that are scalar multiples can be combined. This reversibility motivates the following definition.

\begin{definition}
    Let $A \in \Mr{m}{n}$ be a communication matrix. We say that $A$ is \textit{compressible} if we can form a new matrix $A' \in \Mr{m'}{n'}$ from $A$ by removing or combining any rows or columns, so that $m'<m$ or $n'<n$, and $A \usim A'$. Otherwise $A$ is \textit{incompressible}.
\end{definition}

Clearly the incompressible communication matrices are the smallest members within any ultraweak equivalence class of matrices. 
We observe that an incompressible communication matrix cannot have identical rows or rows contained in the convex hull of other rows. 
An incompressible communication matrix also cannot have zero columns or columns that are scalar multiples of other columns. For any communication matrix $A$, the set $[A]$ will contain incompressible elements.

A measurement device is always normalized in the following sense: we can split existing outcomes into new outcomes probabilistically, but we cannot add arbitrary new outcomes into a measurement device as normalization would be lost. Meanwhile a state ensemble is not restricted in this way; new states can always be added into a state ensemble. Of particular interest to us is the following question: in what situations can the physical action of adding a new state into a state ensemble be simulated via ultraweak matrix majorization? Moreover, if the physical action of adding a new state is simulable via ultraweak matrix majorization, is it sufficient to only modify the state preparation device or is it also necessary to consider non-trivially post-processing the measurement outcomes? In terms of ultraweak matrix majorization, if $A'$ is obtained from $A$ by adding a new state into the state ensemble implementing $A$ and $A \usim A'$, is it sufficient to consider $A' = LA$, or is it also necessary to consider $A' = LAR$? In general, we cannot rule out solutions where $R \neq \id$, as the ultraweak matrix majorization might have multiple solutions. However, we are able to prove a weaker result (Prop. \ref{prop:permutation-prop}) that will be useful later. First we recall a result from an earlier article regarding quantum communication matrices with a maximal value of $\lmaxw$.

\begin{lemma}[\cite{HeKeLe2020}, Prop. 1.]\label{lemma:lmax-lemma}
    Let $A \in \C{m}{n}{\Qd}$ and $\lmax{A} = d$. If $A$ does not have any zero columns, then $A$ has an implementation with a rank-1 POM. The pure eigenstate of each effect must be included in the state ensemble implementing $A$.
\end{lemma}

\begin{proof}
    If $A_{ij} = \tr{\varrho_i \Mo_j}$, then the maximum value in each column of $A$ is limited by the maximum eigenvalue of the corresponding effect, i.e., $A_{ij} \leq r_j$, where $r_j$ is the maximum eigenvalue of $\Mo_j$. The result then follows by observing that \begin{equation}
        \lmax{A} = \sum_j \max_i A_{ij} \leq \sum_j r_j \leq \sum_j \tr{\Mo_j} = \tr{\id} = d,
    \end{equation}where the second inequality is an equality only if $\Mo_j$ is rank-1 (or zero) for all $j$. If $\Mo_j$ is rank-1, then $\max_{\varrho \in \shd} \tr{\varrho \Mo_j} = r_j$ and the maximum is achieved by the pure eigenstate of $\Mo_j$. Thus the state ensemble implementing $A$ must contain the pure eigenstates of the effects of $\Mo$.
\end{proof}

\begin{proposition}\label{prop:permutation-prop}
    Let $A \in \C{m}{n}{\Qd}$ be incompressible and $\lmax{A}=d$. If $B = LAR$ for some $L\in \Mr{m+1}{m}$ and $R\in \Mr{n}{n}$ and, additionally, $\lmax{B}=d$, then $R$ is a permutation matrix.
\end{proposition}
\begin{proof}
    We first note that the assumption of incompressibility does not come at the cost of generality, as there always exists incompressible matrices in the ultraweak equivalence class of any communication matrix. Let us denote the quantum implementation of $A$ as $A_{ij}=  \tr{\varrho_i \Mo_j}$ for some state ensemble $\{\varrho \}_{i=1}^m$ and an $n$-outcome POM $\Mo$. Likewise, let us denote the quantum implementation of $B$ as $B_{ij}=  \tr{\varrho'_i \Mo'_j}$ for a state ensemble $\{\varrho' \}_{i=1}^{m+1}$ and an $n$-outcome POM $\Mo'$.
    
    Since $\lmax{A}=d$, we know that $\Mo$ is rank-1 and $\{\varrho \}_{i=1}^m$ contains the pure eigenstates of the effects of $\Mo$ by Lemma \ref{lemma:lmax-lemma}. There are no zero effects since $A$ is incompressible. We also note that $A$ has no duplicate rows, meaning that all states in $\{\varrho \}_{i=1}^m$ are different, and that none of the effects in the POM are scalar multiples of other effects. If this was the case we could remove the duplicate rows and combine the effects that are proportional without affecting the ultraweak equivalence class.

    Now, since $B=LAR$, $\lmax{B}=d$ and $B$ contains the same number of columns as $A$, we know that $\Mo'$ is rank-1 and the corresponding eigenstates are included in the state ensemble $\{\varrho' \}_{i=1}^{m+1}$. Based on ultraweak majorization $B$ has an implementation with the states $\{\varrho' \}_{i=1}^{m+1} = \{ \sum_{k=1}^m L_{ik} \varrho_k \}_{i=1}^{m+1}$ and effects $\{ \Mo'_j\}_{j=1}^n =  \{ \sum_{k=1}^n R_{kj} \Mo_k \}_{j=1}^{n}$. As each of the effects of $\Mo$ and $\Mo'$ are rank-1 and none of the effects of $\Mo$ are proportional to other effects of $\Mo$, we conclude that $R$ must be a permutation matrix, as otherwise $\Mo'$ would contain an effect that is not rank-1.
\end{proof}

Based on Prop. \ref{prop:permutation-prop} we are not exactly able to resolve the question of simulating the action of adding a new state into a state ensemble via ultraweak matrix majorization. However, it turns out this result is just strong enough for the characterization of maximal communication matrices in the next section.  We remark that if $A \in \Call{\Qd}$ is incompressible, $\lmax{A}=d$, $B$ is obtained by adding a new state into the implementation of $A$ and $B \usim A$, then $[ B ]$ contains a matrix $B'$ such that $B' = LA$ for some row-stochastic matrix $L$ because permutation matrices are invertible and the inverse of a permutation matrix is another permutation matrix. The significance of the equation $B' = LA$ is that the equivalence class $[ B ]$ should contain a matrix whose rows are contained in the convex hull of the rows of $A$ \cite[Prop. 3.3]{Martinez2005Weak}.


We are now equipped to prove our main results which are presented in the next section.

\section{Maximal quantum communication matrices}\label{section:maximal-communication-matrices}

An interesting and mathematically basic question is whether a partially ordered set contains the greatest and the least elements. 
In the context of communication matrices and ultraweak majorization, this corresponds to the question of characterizing communication setups that are easiest and hardest to implement.
As we explained earlier, a set of classical communication matrices has the greatest element $[\id_d]$ for $d$-dimensional classical theory. It is not difficult to see that any theory has the equivalence class of the communication matrix $V_n \in \Mr{m}{n}$, $(V_n)_{ab}=\frac{1}{n}$ for all $a,b$ as the least element. This is because the communication matrix $V_n$ is easily seen to be majorized by all other communication matrices. Operationally we can interpret $V_n$ as a matrix of total uniform noise.

We conclude that every $A \in \Call{\Cd}$ satisfies $V_n \uleq A \uleq \id_d$. 
This motivates us to define the compatibility of communication matrices.

\begin{definition}\label{def:compatibility-of-matrices}
    A communication matrix $A \in \Call{\state}$ is $\Call{\state}$-compatible with another communication matrix $B \in \Call{\state}$ if and only if there exists a third communication matrix $C \in \Call{\state}$ such that $A \uleq C$ and $B \uleq C$.
\end{definition}

The prefix in $\Call{\state}$-compatible can be dropped when there is no risk of confusion. Operationally Def. \ref{def:compatibility-of-matrices} means that for compatible communication matrices there exist state preparation and measurement devices that are able to implement both matrices with the aid of pre- and post-processing. 
As all classical communication matrices are below the identity it follows that all classical communication matrices are compatible. 

Although all communication matrices are seen to be compatible in classical theory, a nontrivial question is computing the least upper bound of two communication matrices. This is left for future work. Instead we will focus on characterising the maximal communication matrices in quantum theory. This is an important first step to studying the question of compatibility.

So far we have not seen any evidence, in the current or previous works, that quantum theory should contain maximal elements in the partially ordered set of communication matrices. Indeed, a partially ordered set need not have maximal or minimal elements in general.  As we do not yet know if any quantum communication matrix meets the criteria of maximality, we should be wary of assuming such matrices exist. However, we can prove the following requirements for a quantum communication matrix to be maximal.

\begin{lemma}
    Let $A \in \C{m}{n}{\Qd}$ be ultraweakly maximal. Then $\lmax{A} = d$ and $\lmin{A} = 0$.
\end{lemma}
\begin{proof}
    Without loss of generality we assume that $A$ does not contain any zero columns, i.e., none of the effects in its implementation are zero. Let us first consider $\lmaxw$. As $A \in \C{m}{n}{\Qd}$ there exists a state ensemble $\{ \varrho_a \}_{a=1}^m$ in $\mathcal{S}(\mathcal{H}_d)$ and a POM $\Mo = \{ \Mo_b \}_{b=1}^n$, such that $A_{ab} = \tr{\varrho_a \Mo_b}$. If $\Mo$ contains any decomposable effects we can form a new communication matrix $A'$ from $A$ where all decomposable effects have been decomposed into rank-1 effects. Clearly $A \uleq A'$. If the state preparation device used to implement $A'$ is not able to prepare the pure eigenstates of the decomposed POM, we can add these eigenstates to the state preparation device, forming a new communication matrix $A''$. Clearly $A' \uleq A''$. It is now evident that $\lmax{A''} = d$ and as we assumed $A$ to be maximal it follows that $\lmax{A}=d$ as $A \usim A''$.

    We have now proved that if $A$ is maximal then the measurement device implementing $A$ consists of rank-1 effects and $\lmax{A}=d$. Each rank-1 effect has a nontrivial kernel. Based on the dimension of the quantum implementation, for each effect $\Mo_x$ there exists at least one pure state $\varrho_x$ such that $\tr{\varrho_x \Mo_x}=0$. If the preparation device implementing $A$ is not able to prepare the states $\varrho_x$, we can add one such state for each effect into the device forming a new communication matrix $A^*$. Clearly $A \uleq A^*$. Now we have that $\lmin{A^*}=0$, and as $A \usim A^*$ as $A$ was assumed maximal, it follows that $\lmin{A}=0$.
\end{proof}

We have not yet proved that maximal elements exist in $\Call{\Qd}$. Let us now prove that at least one maximal element exists.

\begin{proposition}\label{prop:id-is-maximal}
    Let $A \in \C{m}{n}{\Qd}$. If $\id_d \uleq A$, then $A \usim \id_d$, i.e., $[\id_d]$ is maximal in $\Calle{\Qd}$.
\end{proposition}
\begin{proof}
    Let us first recall what we know about $\id_d$. We have that $\lmax{\id_d}=d$, $\lmin{\id_d}=0$ and $\iota(\id_d)=d$. These are the maximal values of the monotones in $\Call{\Qd}$. 
    Assuming $\id_d \uleq A$, it follows that $A$ must have these maximal values too. 
    Without loss of generality let us assume that $A$ is incompressible. From $\lmax{A}=d$ we know that the POM implementing $A$ is rank-1 and the state preparation device is able to prepare the pure eigenstates of the POM. We also know from $\iota(A)=d$ that $A$ must have $d$ mutually orthogonal rows \cite[Proposition 9]{HeKeLe2020}. 

    Suppose $A$ is implementable with the state ensemble $\{ \varrho_i \}_{i=1}^m$ and an $n$-outcome POM $\Mo$. Without loss of generality let the first $d$ states ($A$ has at least $d$ rows and columns, otherwise $\lmax{A}<d$) be the ones that correspond to orthogonal rows in the communication matrix. We observe that the support of each effect on this set is disjoint. Let $\text{supp}\left( \varrho_k \right) = \{ j \in \{ n\}\, | \, \tr{\rho_k \Mo_j} > 0 \}$ for $k\in \{ d \}$. Then we have \begin{equation}\label{eq:state-support}
        \sum_{j \in \text{supp}\left( \varrho_k \right) } \tr{\varrho_k \Mo_j} = 1.
    \end{equation}
    Each of the effects in \eqref{eq:state-support} is a rank-1 operator. Let us denote with $\No_k = \sum_{j \in \text{supp}\left( \varrho_k \right) } \Mo_j$. Now the communication matrix $B\in \C{d}{d}{\Qd}$ defined by $B_{ab} = \tr{\varrho_a \No_b}$ clearly equals $\id_d$ (or a permutation of $\id_d$). It follows from Lemma \ref{lemma:lmax-lemma} that $\No_k$ is a rank-1 operator for all $k$ and $\varrho_k$ is the unique eigenstate of $\No_k$, i.e. a pure state. Clearly now each of the effects in the sum of $\No_k$ must be a scalar multiple of $\No_k$ itself. This is contrary to our earlier assumption that $A$ is incompressible. As $\tr{\varrho_k \No_k} = 1$ and $\No_k$ is rank-1, it follows that $\No_k$ must be a projection for all $k$. Finally, as $\sum_k \No_k = \id_d$, we have that $\No_i \No_j = 0$ for all $i \neq j$, i.e., the effects of $\No$ are pairwise orthogonal.

    So far we have proved that if $\id_d \uleq A$ and $A$ is incompressible, then $A$ necessarily has an implementation with a sharp rank-1 POM. As the first $d$ rows of $A$ were orthogonal, we have that the first $d$ rows of $A$ actually form (a permutation of) the identity matrix $\id_d$. But $A$ can also have other rows. The proof is completed by observing that a rank-1 sharp POM cannot distinguish any state $\varrho_x \in \shd$ from the mixture $\varrho'_x=\sum_i \tr{\varrho_x \Mo_i}P_i$, where $P_i$ is the eigenstate of the rank-1 $\Mo_i$. Thus whatever states $\varrho_x$ were included in the implementation of $A$, we can simply add the mixtures $\varrho'_x$ to the implementation of $\id_d$ instead. This shows that $\id_d \usim A$.
\end{proof}

\begin{figure}
    \centering
    \includegraphics[width=\textwidth]{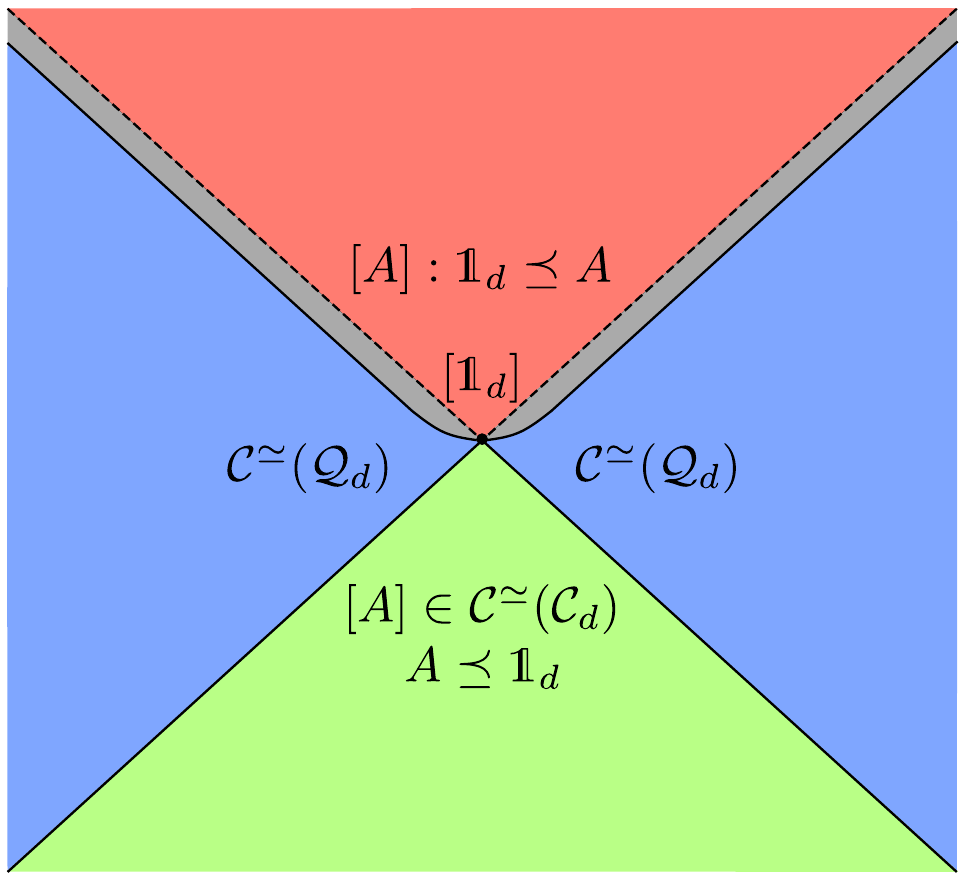}
    \caption{Depiction of the partially ordered set of (equivalence classes of) communication matrices. The center point represents the ultraweak equivalence class of $\id_d$. The upper red triangle depicts communication matrices that ultraweakly majorize $\id_d$. Similarly the lower green triangle depicts those matrices that are majorized by $\id_d$. The blue areas represent equivalence classes in $\Calle{\Qd}$. As no matrix in $\Call{\Qd}$ is stronger than $\id_d$, no matrix in $\Call{\Qd}$ can lie on the dashed line on the boundary of the red triangle. This fact is illustrated by the gray area between the red and blue sets.}
    \label{fig:preprder-structure}
\end{figure}

The sets of classical and quantum communication matrices can now be seen to share some structure; the identity matrix is a maximal element in both of them. 
We have seen that the identity is the greatest element in classical theory. 
With the following lemma we will prove our main result, namely that the identity is the only maximal element in quantum theory, although not the greatest. 
Our main result is illustrated in Fig. \ref{fig:preprder-structure}, which depicts the partially ordered set of communication matrices in light of Thm. \ref{thm:main-result}.

\begin{lemma}\label{lemma:projection-lemma}
    Let $E=P_1 + t P_2$, where $P_1 = \kb{\varphi}{\varphi}$, $P_2 = \kb{\psi}{\psi}$, $P_1 \neq P_2$ are non-orthogonal 1-dimensional projections, $t \in \mathbb{R}$, $t \neq 0$, and $\varphi, \psi \in \hi$ for some complex separable Hilbert space $\hi$. The operator $E$ has two eigenvalues
    $$
\lambda^{\pm}_t = \frac{1+t}{2}\pm \sqrt{\frac{(1+t)^2}{4} - t(1- |z|^2)} \, ,
    $$
    where $z = \ip{\varphi}{\psi}$.
    An unnormalized eigenvector corresponding to the largest eigenvalue $\lambda^+_t$ is $z \varphi + (\lambda^+_t -1) \psi$.
\end{lemma}

\begin{proof}
    Let us calculate the eigenvalues of $E$ explicitly. Let $\phi \in \hi$. Then \begin{align}
        E \phi = (P_1 + tP_2)\phi = \ip{\varphi}{\phi}\varphi + t \ip{\psi}{\phi}\psi.
    \end{align}
    This shows that the eigenvectors of $E$ are of the form $c_1 \varphi + c_2 \psi$ for some complex numbers $c_1, c_2 \in \mathbb{C}$. Inserting $\phi = c_1 \varphi + c_2 \psi$ into $E\phi = \lambda \phi$, we obtain \begin{align}
        \lambda c_1 &= c_1 + c_2 \ip{\varphi}{\psi} = c_1 + c_2 z \label{eq:eigenvector-1}\\
        \lambda c_2 &= t \left( c_1 \ip{\psi}{\varphi} + c_2 \right) = t \left( c_1 \bar{z} +  c_2 \right),\label{eq:eigenvector-2}
    \end{align} where $z = \ip{\varphi}{\psi}$. Eliminating the variable $\frac{c_2}{c_1}=c$ yields \begin{align*}
        \lambda^{\pm}_t = \frac{1+t}{2}\pm \sqrt{\frac{(1+t)^2}{4} - t(1- |z|^2)}.
    \end{align*} If $\lambda^+_t = \lambda^-_t$, then the term inside the square root in the above equation must equal zero. Solving for $t$ in $\frac{(1+t)^2}{4} - t(1- |z|^2)=0$ and requiring $t$ to be real ultimately yields $|z|^4 - |z|^2 \geq 0$ which only holds for $|z|^2=0$ or $|z|^2=1$ as generally $0 \leq |z|^2 \leq 1$. 

    Finally, setting $c_1 = z$ in Eq. \eqref{eq:eigenvector-1}, we get $c_2 = \lambda  - 1$, so an unnormalized eigenvector corresponding to the eigenvalue $\lambda_t^+$ is $z \varphi + (\lambda_t^+ -1)\psi$. We observe that $\lambda_t^+ = 1$ if and only if $t=0$ and $\lim_{t \rightarrow \infty} \lambda_t^+ = \infty$, $\lim_{t \rightarrow -\infty} \lambda_t^+ = 1 - |z|^2$. Notably an eigenvector of $E$ corresponding to $\lambda_t^+$ is different from $\varphi$ or $\psi$ whenever $t \neq 0$.
\end{proof}

\begin{theorem}\label{thm:main-result}
    Let $A \in \C{m}{n}{\Qd}$. If $A$ is an ultraweakly maximal element in $\Call{\Qd}$, then $A \usim \id_d$, i.e., $[\id_d]$ is the only maximal element in $\Calle{\Qd}$.
\end{theorem}
\begin{proof}
    Suppose $A\in \Call{\Qd}$ is maximal. Then $\lmax{A}=d$ and $A$ has an implementation $A_{ab}=\tr{\varrho_a \Mo_b}$, where $\Mo_b$'s are either rank-1 or rank zero and the state ensemble $\{ \varrho_i \}_{i=1}^m$ contains the eigenstates of the non-zero effects of $\Mo$ by Lemma \ref{lemma:lmax-lemma}. Observe that if the effects of $\Mo$ are pairwise orthogonal, then $\Mo$ is post-processing equivalent with a sharp rank-1 POM, and in this case $A \usim \id_d$ by Prop. \ref{prop:id-is-maximal}. Without loss of generality let us assume $A$ is incompressible.

    Assume then that there exists a pair of effects, $\Mo_i$ and $\Mo_j$, such that $\tr{\Mo_i \Mo_j} \neq 0$. Consider the following post-processing of these effects: $\Mo_k = a\Mo_i + b \Mo_j$, $0 < a,b < 1$. We do not need to create the new effect $\Mo_k$ via post-processing as we are able to calculate $\tr{\varrho_i \Mo_k}$ directly from the communication matrix for any $i$. By Lemma \ref{lemma:projection-lemma} the positive operator $\Mo_k$ has a singular maximal eigenvalue whenever $a,b \neq 0$. The eigenstate corresponding to this eigenvalue is a superposition of the eigenstates of $\Mo_i$ and $\Mo_j$. Let us denote this state by $\varrho_k$. In particular, as $\varrho_k$ is distinct from the eigenstates of $\Mo_i$ and $\Mo_j$ and approaches them as $a \rightarrow 0 $ or $b \rightarrow 0$, we conclude that there is an infinite number of possible eigenstates of $\Mo_k$ corresponding to different values of $a$ and $b$.

    Let us now fix the values of $a$ and $b$, $a,b\neq 0$. As $A$ was assumed maximal, we must have $A \usim A'$, where $A'$ is formed from $A$ by adding $\varrho_k$ to the state ensemble $\{ \varrho_i \}_{i=1}^m$. We can always choose such $a$ and $b$ that $\varrho_k$ is not already contained in the state ensemble.  Note that it is not necessary for $\Mo$ to contain $\Mo_k$ as an effect. However, we have that $\mathrm{max}_{\varrho \in \shd} \tr{\varrho \Mo_k} = \tr{\varrho_k \Mo_k}$ is a unique solution. In particular this means that the row corresponding to $\varrho_k$ in $A'$ cannot be a convex combination of rows in $A$.
    
    The proof is now concluded by observing that $\lmax{A}=d$ and $A$ was incompressible. Therefore, if $A \usim A'$, then $A'$ has an implementation with the state ensemble $\{ \varrho'_i \}_{i=1}^{m+1} = \{ \sum_{k=1}^{m} L_{ik} \varrho_k \}_{i=1}^{m+1}$ and a POM with the effects $\{ \Mo'_j \}_{j=1}^n = \{ \sum_{k=1}^n R_{kj} \Mo_k \}_{j=1}^n$ for some row-stochastic matrices $L \in \Mr{m+1}{m}$ and $R\in \Mr{n}{n}$ based on ultraweak majorization. Moreover, by Prop. \ref{prop:permutation-prop}, $R$ must be a permutation matrix. Hence the POM $\Mo'$ still contains the effects $\Mo_i$ and $\Mo_j$ although the outcomes might have been relabelled. Suppose these effects correspond to $\Mo'_{i'}$ and $\Mo'_{j'}$ so that $\Mo_k = a\Mo'_{i'} + b\Mo'_{j'}$. However, as we already concluded, $\mathrm{max}_{\varrho \in \shd} \tr{\varrho \Mo_k} = \tr{\varrho_k \Mo_k}$ is a unique solution where $\varrho_k$ is the pure eigenstate of $\Mo_k$. Therefore it is not possible that $\tr{\varrho'_{m+1}\left( a\Mo'_{i'} + b\Mo'_{j'} \right)} = \tr{\varrho_k \Mo_k}$ as $\varrho_k$ was not contained in the ensemble $\{ \varrho_i \}_{i=1}^m$. This is a contradiction with the assumption that $A$ is maximal. Hence, it follows that the POM implementing $A$ is rank-1 and does not contain non-orthogonal effects, which by Prop. \ref{prop:id-is-maximal} implies $A \usim \id_d$.
\end{proof}

\section{Discussion}\label{section:discussion}

Our main result showed that $[\id_d]$ is the only maximal element in $\Calle{\Qd}$. The technique we used in our proof gives a mathematically intuitive reason for why this holds. Unless the quantum implementation of a communication matrix is done with a rank-1 sharp POM, we are always able to ``strengthen'' a communication matrix by adding new pure states into the state ensemble implementing it as long as there are at least some rank-1 effects. As there are infinitely many pure states in quantum theory of any dimension, we will never run out of possible new states to add. We thus run into these sequences of communication matrices $A \uleq A' \uleq A'' \uleq \dots$, where each communication matrix is obtained from the previous by adding a state and each communication matrix in the sequence is genuinely stronger than the previous.

Although we are able to create these growing sequences of communication matrices, no quantum communication matrix is stronger than the identity. This result is surprising, yet in good agreement with Holevo's theorem \cite{Holevo1973} which states that the information transmitted by a quantum system cannot exceed its operational dimension. It is important to stress that the set $\Call{\Qd}$ is not identical to $\Call{\Cd}$. 
In particular, $\Call{\Qd}$ contains communication matrices that are ultraweakly incomparable with $\id_d$ (Example \ref{ex:incomparable}).
As $[\id_d]$ is maximal it now follows that there must exist sets of incompatible communication matrices in $\Call{\Qd}$.

Finally, we note that the question of maximal communication matrices can be studied in the framework of general probabilistic theories \cite{Plavala2023General}. An open question is if classical and quantum theories are the only operational theories where the only maximal element is the identity. We leave this question, together with the question of incompatibility, for future works.

\section*{Acknowledgments}


TH and OK acknowledge financial support from the Business Finland under the project TORQS, Grant 8582/31/2022, and from the Academy of Finland under the mobility funding Grant No. 343228, and under the project DEQSE, Grant No. 349945

\bibliographystyle{quantum}
\bibliography{bibliography.bib}

\end{document}